\documentstyle[12pt,epsfig]{article}
\begin{document}
\vspace{1.0in}
\begin{center}
{\large\bf On the microscopic basis of Newton's law of cooling and
beyond}
\end{center}
\begin{center}
\vspace{1cm} {\bf Mihir Ranjan Nath {\footnote[1]
{mrnath\_95@rediffmail.com}} and Surajit Sen {\footnote[2]
{ssen55@yahoo.com}}\\
Department of Physics\\
Guru Charan College\\Silchar 788004, INDIA\\}
\vspace {1cm}
%{\bf Surajit Sen \footnote[2] {ssen55@yahoo.com}\\
%Department of Physics\\
%Guru Charan College\\
%Silchar 788004, INDIA} \\
\vspace {1cm}
{\bf Gautam Gangopadhyay \footnote[3] {gautam@bose.res.in}\\
S N Bose National Centre for Basic Sciences\\
JD Block, Sector III, Salt Lake City \\
Kolkata 700098, INDIA} \\
\vspace {.5cm}
\end{center}
%\vfill
%\pagebreak
\begin{abstract}
The microscopic basis of Newton's law of cooling and its
modification when the difference in temperature between the system
and the surroundings is very large is discussed. When the system of
interest is interacting with a small bath, the effect of the
dynamical evolution of the bath variables is important to find out
its dynamical feedback on the system. As in the usual system-bath
approach, however, the bath is finally considered to be in thermal
equilibrium and thereby provides an effective generalization of the
Born-Markov master equation. It is shown that the cooling at early
time is faster than that predicted by Newton's law due to the
dynamical feedback of the bath.
\end{abstract}
%\vspace{1in}
\vfill \pagebreak
\begin{center}
\large I.Introduction
\end{center}
\par
Classically the phenomenon of cooling of a bulk body may be
considered as a process where the flow of heat from the hotter body
to a colder environment is governed by the Newton's law of cooling,
namely,
\begin{flushleft}
\hspace{2in} $-\frac{dT(t)}{dt}=\gamma(T(t)-T_R),$\hfill (1)
\end{flushleft}
with $T(t)$ be the instantaneous temperature of the body, $T_R$ be the
temperature of the environment and $\gamma$ be the characteristic decay
constant, respectively. The solution of Eq.(1) reads
\begin{flushleft}
\hspace{2in} $T(t)=T_0e^{-\gamma t}+ T_R(1-e^{-\gamma t})$,\hfill(2)
\end{flushleft}
where $T_0$ is the initial temperature of the body at $t=0$ such
that $T_R<T_0$. The classical rate equation does not contain any
non-linear term and it is valid if the difference between the
temperature of the system and environment is small.
\par
On the other hand, the cooling of the neutral atoms is generally
performed by various coherent laser-cooling techniques, which
primarily concern with the reduction of the kinetic motion of the
center-of-mass of the trapped atoms [1,2]. However, if we consider
the coherent cooling of the molecules [3], then the contribution to
the thermal energy arises not only from the motion of their centre
of mass, but also from their rotational or vibrational motion. Thus
for an ensemble of trapped molecules the cooling requires the
ceasing of the momentum associated with all possible degrees of
freedom, although a priori it is difficult to ascertain which of the
degree of freedom contributes most significantly. If we neglect the
translational and rotational motion of the trapped molecules, the
dissipation of temperature associated with the vibrational degrees
of freedom may be formulated quantitatively within the framework of
the standard dissipation theory of the damped harmonic oscillator.
To formulate the theoretical basis of the vibrational cooling, we
assume that a molecule with one or few modes of vibration as our
system of interest which dissipates its energy into the large number
of other modes acting effectively acts as the reservoir. In the
density matrix formalism of the system-reservoir composite model,
the nontrivial weak coupling between the system with the reservoir
effectively induces a damping in the system. Consequently, we obtain
the Born-Markov master equation of the reduced density matrix
$\rho$, where the reservoir oscillators are completely eliminated in
terms of the system variables [4,5] as
\begin{flushleft}
\hspace{2in} $\frac{d\rho}{dt}=-i\omega_0 [a^\dagger a,\rho] -
\gamma(1+\bar{n}_R(T_R))( a^\dagger a \rho - 2a\rho a^\dagger+ \rho
a^\dagger a) -\gamma\bar{n}_R(T_R)( a a^\dagger \rho -2a^\dagger
\rho a+\rho a a^\dagger), $\hfill (3)
\end{flushleft}
where the frequency of the system oscillator is $\omega_0$ with $[a,
a^\dagger] =1$ and the average thermal excitation number of the bath
is $\displaystyle \bar{n}_R(T_R) =\frac{1}{e^{\frac{\hbar\omega_0}{k
T_R}}-1}$, where $T_R$ is the temperature of the reservoir. Thus, it
follows from the master equation that the time evolution of average
photon number of the system mode, $\frac{d\bar{n}_S(t)}{dt}=
Tr[a^\dagger a \frac{d\rho}{dt}]$ with frequency $\omega_0$ is
described by the rate equation
\begin{flushleft}
\hspace{2in} $\displaystyle -\frac{d\bar
n_S(t)}{dt}=\gamma(\bar{n}_S(t)-\bar{n}_R(T_R)),$\hfill (4)
\end{flushleft}
and its solution is given by
\begin{flushleft}
\hspace{2in} $\displaystyle \bar{n}_S(t) = \bar{n}_R(T_R)
+(\bar{n}_S(0)- \bar{n}_R(T_R)) \displaystyle e^{-\gamma t}.$
\hfill(5)
%\hspace{2in} $\bar{n}_S(t) =\bar{n}_S(0)e^{-\gamma t}
%+\bar{n}_R(T_R)(1-e^{-\gamma t}).$
\end{flushleft}
Thus, the vibrational cooling corresponds to the feeding of the
thermal photon from the system to the reservoir until the system
photon number equilibrates with that of the reservoir. This process
is known as the thermalization and it is evident from Eq.(5) that,
similar to the classical cooling, it occurs after an infinitely
large time, namely, $\bar{n}_S(\infty)=\bar{n}_R(T_R)$. We can
associate the instantaneous average photon number of the system mode
with an effective temperature $T(t)$ by using the relation
\begin{flushleft}
\hspace{1.5in}
$\displaystyle \bar{n}_S(T(t)) =
\frac{1}{e^{\frac{\hbar\omega_0}{kT(t)}}-1}$,
\hfill (6)
\end{flushleft}
which in the high temperature limit gives, $\bar{n}_S(T(t)) \approx
\frac{kT(t)}{\hbar\omega_0}$. In the same limit, the initial system
photon number and the reservoir photon number are given by
$\bar{n}_S(T_0) \approx \frac{kT_0}{\hbar\omega_0}$ and
$\bar{n}_R(T_R) \approx \frac{kT_R}{\hbar\omega_0}$, respectively;
putting these values in Eq.(5) we recover Eq.(2). Therefore, the
classical Newton's cooling law in Eqs.(1) and (2) can be obtained
from the high temperature limiting situation of the Born-Markov
master equation in Eqs.(3) and (5). This is valid when the
difference in average energy per mode between the system and the
reservoir is small, whereby the reservoir does not change with time
due to the acceptance of energy from the system of interest.
\par
In the crossroad of various approaches and applications of the
system-reservoir composite formalism, the necessity of the finite
bandwidth of the reservoir [6-12] was envisaged right from the
beginning which leads to the possible modification of Eq.(3). The
recent experiments involving ultra-fast time scale [13-15],
correlated emission laser (CEL) pulse with adjustable memory time
[16], experiments on cavity electrodynamics [17] etc, have
significant impact in the understanding of the models beyond the
Born-Markov approximation. However, in all previous studies, the
assumption that works at more subtle level is that the photon
absorbed by the reservoir from the system cannot bring any dynamical
change with it because of the small difference of the energy between
the system and average energy of each degree of freedom of the
reservoir. It is therefore of natural interest how the situation
changes if we consider that the energy of the system is large enough
in comparison to that of  each mode of the reservoir. In this paper
we shall show that, due to the large difference of average energy
between the system and each reservoir mode, the flow of the thermal
photon of substantial energy from the system to the reservoir
effectively leads to the dynamical evolution of the reservoir. Our
primary objective is to discuss the vibrational cooling of a system
with large energy content by incorporating the aforesaid dynamical
evolution of the reservoir.
\par
The remaining sections of the paper are organized as follows. In
Sec. II we have developed a formalism to incorporate the evolution
of the reservoir and show how it effectively generalizes the
Born-Markov master equation beyond leading order of the decay
constant. We apply our formalism to the damped harmonic oscillator
in Sec. III and show that how it affects the thermalization time to
make it short. We conclude by highlighting the essence of the paper
and discussing its outlook.
\begin{center}
\large II. Formalism
\end{center}
\par
The Hamiltonian of a system interacting with the reservoir is
given by
\begin{flushleft}
\hspace{1in} $H = H_S + H_R +V \equiv H_0 + V,$\hfill(7)
\end{flushleft}
where $H_S$, $H_R$ represent the Hamiltonians of the system and
reservoir, respectively and $V$ is the interaction between them. Let
$\kappa(t)$ be the joint density matrix of the composite system in
the Interaction Picture (IP); the corresponding evolution equation
is
\begin{flushleft}
\hspace {1in}
 $\displaystyle
\frac{{\partial \kappa(t)}}{{\partial t}} = - \frac{i}{\hbar}[V(t),
\kappa(t)].$\hfill(8)
\end{flushleft}
The solution of the equation is given by
\begin{flushleft}
\hspace {1in}
 $\kappa (t) = \kappa (0) - \frac{i}{{\hbar
}}\int\limits_0^t {dt'} [V(t'), \kappa (0)] + (\frac{-i}{\hbar
})^2 \int\limits_0^t {dt} '\int\limits_0^{t'} {dt''} [V(t'),
[V(t''), \kappa(t'')]] $.\hfill(9)
\end{flushleft}
We consider the interaction Hamiltonian in the IP is of the
following form:
\begin{flushleft}
\hspace {1.0in} $V(t) =\hbar\sum\limits_i{Q_i(t)F_i(t)}$,\hfill(10)
\end{flushleft}
where $Q_i(t)$ and $F_i(t)$ be the system and reservoir operators,
respectively, in IP. Using the factorization ansatz, namely,
$\kappa(0)\approx s(0) f(0)$ and $\kappa(t'')\approx s(t'') f(t'')$,
respectively, and by noting the fact that $Tr_R\kappa(t)= s(t)$, the
trace over the reservoir mode in Eq.(9) yields
\begin{flushleft}
\hspace{1in}
%\begin{array}{l}
$ s(t) = s(0) -
i \int\limits_0^t {dt'} \sum\limits_i  \langle F_i (t')\rangle_R
[Q_i(t'),  s(0)]\  - $  \\
\hspace{1in} \vspace{.4cm} $\int\limits_0^t {dt'}
\int\limits_0^{t'} {dt''} \sum\limits_{i,j}\{{
Tr_R[F_i(t')F_j(t'') f(t'')]
%\{ W_{ij} ^+[t,t']
[Q_i(t')Q_j(t'')s(t'') - Q_j(t'')s(t'')Q_i(t')] -} $ \\
\hspace{1in} \vspace{.4cm} $Tr_R[F_j(t'')F_i(t')f(t'')]
%W_{ji}^ -[t,t']
[Q_i(t')s(t'')Q_j(t'') - s(t'')Q_j(t'')Q_i(t')]\},$\hfill(11)
%\end{array}
\end{flushleft}
where $\langle {...}\rangle_R = Tr_R[...f(0)]$ is the average of the
reservoir operators. Taking the derivative of Eq.(11) with respect
to $t'$ we obtain (we redefine $t''$ by $t'$ and $t'$ by $t$),
\begin{flushleft}
\hspace{1in} $\displaystyle \frac{{\partial s(t)}}{{\partial t}} = -
\sum\limits_{i,j} \int\limits_0^{t} {dt'} {\{[Q_i Q_j s(t') -
Q_js(t')Q_i]
Tr_R[F_i(t)F_j(t') f(t')] - } $ \\
\hspace{1in} \vspace{.4cm} $[Q_i s(t')Q_j- s(t')Q_j Q_i]
Tr_R[F_j(t')F_i(t)
f(t')]\}exp[i(\omega_i^St+\omega_j^St'),$\hfill(12)
\end{flushleft}
where $\omega_i^S$ be the characteristic frequencies of the system
and the term linear in reservoir operator vanishes by the symmetry
argument. The system oscillator in the IP in V(t) can be expressed
in the Schrodinger picture (SP) by using the standard prescription,
\begin{flushleft}
\hspace{1in} $Q_i(t)=e^{(i/\hbar) H_S t} Q_i
e^{-(i/\hbar)(H_S)t}$
\end{flushleft}
\begin{flushleft}
\hspace{1.4in} $= Q_i e^{-i\omega_i^S t}$.\hfill(13)
\end{flushleft}
Now, replacing $t'$ by $t-\tau$ in Eq.(12) and assuming the
Born-Markov approximation [4], namely, $s(t-\tau)\approx s(t)$ for
large t, the generalized master equation of the reduced density
operator in Schrodinger picture, S, is obtained as
\begin{flushleft}
\hspace{1in} $\frac{d S}{dt}=-\frac{i}{\hbar}[H_S, S]-
%\int\limits_0^t {dt'}
\sum\limits_{i,j}{\{ [Q_i Q_j S - Q_j S Q_i]W_{ij} ^ +
[t] - } $ \\
\hspace{1.5in} \vspace{.2cm} $[Q_i S Q_j - S Q_j Q_i]W_{ji}^
-[t]\} \delta(\omega_i^S+\omega_j^S),$\hfill(14)
\end{flushleft}
where
\begin{flushleft}
\hspace{1in}
 $W_{ij}^+[t]=\int\limits_0^{t}{d\tau}e^{i\omega_i^S\tau}
Tr_R[F_i(t)F_j(t-\tau) f(t-\tau)],$\hfill(15a)
\end{flushleft}
\begin{flushleft}
\hspace{1in}
 $W_{ji}^-[t]=\int\limits_0^{t}{d\tau}e^{i\omega_i^S\tau}
Tr_R[F_j(t-\tau)F_i(t) f(t-\tau)],$\hfill(15b)
\end{flushleft}
which are to be calculated in different situations.
\par
To include the evolution of the reservoir in this scenario, using
$Tr_S\kappa(t)= f(t)$  along with
Eq.(10), the trace of Eq.(9) over the system yields
%\begin{array}{l}
\begin{flushleft}
%\begin{array}{l}
\hspace{1.0in}
 $f(t-\tau) = f(0) -i \sum\limits_k
{\int\limits_0^{t-\tau} {dt_1 }\langle }
Q_k (t_1)\rangle _S [F_k (t_1 ), f(0)] - $ \\
\hspace{1in} \vspace{.2cm}
$\sum\limits_{l,m}{\int\limits_0^{t-\tau} {dt_1 } } \int\limits_0^{t_1
} {dt_2 } [(F_l (t_1 )F_m (t_2 )f(0) - F_m (t_2 )f(0)F_l (t_1 ))\langle
Q_l(t_1)Q_m(t_2)\rangle _S  - $ \\
\hspace{1in} \vspace{.2cm} $ (F_l (t_1 )f(0)F_m (t_2 )-f(0)F_m
(t_2 )F_l (t_1 ))\langle Q_m(t_2) Q_l(t_1) \rangle _S]+...,
$\hfill(16)
 %\end{array}\]
\end{flushleft}
where $\langle...\rangle_S=Tr_S[...s(0)]$ represents the average of
the system operator which depends on the initial population
distribution of the system. In deriving Eq.(16), unlike previous
case, we use the ansatz $\kappa(t'') \approx \kappa(0)$ to terminate
the series beyond the second order in interaction [4]. Plucking
Eq.(16) back into Eq.(15a), we find
\begin{flushleft}
\hspace{1.0in}
%\begin{array}{l}
$W_{ij}^+[t] =W_{ij}^{+0}[t] +\widehat{W}_{ij}^+[t]+...,
$\hfill(17)
\end{flushleft}
where
$W_{ij}^{+0}[t]=\int\limits_0^{t}{d\tau}e^{i\omega_i^S\tau}\langle
F_i(t) F_j(t-\tau)\rangle_R$  is the usual lowest order reservoir
correlator [4,5]. In Eq.(17) the term next to lowest order arises
due to the correlation among the system oscillators, and it is given
by
\begin{flushleft}
\hspace{.5in}
 $\widehat{W}_{ij}^+[t]=
-i \sum\limits_k \int\limits_0^{t} d\tau
e^{i\omega_i^S\tau}{\int\limits_0^{t-\tau} dt_1{ \langle } } Q_k (t_1)\rangle _S \{ \langle F_i(t)
F_j(t-\tau) F_k(t_1) \rangle _R  - \langle F_k(t_1) F_i (t)F_j(t-\tau)
\rangle _R \}  - $ \\
\hspace{0in} \vspace{.4cm}
$\sum\limits_{l,m} \int\limits_0^{t}
{d\tau}e^{i\omega_i^S\tau}{\int\limits_0^{t-\tau} {dt_1 } }
\int\limits_0^{t_1 } {dt_2 }$ \hspace{1in} \vspace{.4cm}
 $ [(\langle F_i(t) F_j(t-\tau) F_l(t_1) F_m(t_2) \rangle _R  -
 \langle F_l(t_1)F_i(t) F_j(t-\tau) F_m(t_2) \rangle _R )\langle
Q_l(t_1) Q_m(t_2) \rangle _S  + $ \\
\hspace{0in} \vspace{.4cm}
 $(\langle F_m(t_2) F_l(t_1) F_i(t) F_j(t-\tau) \rangle _R  - \langle
F_m(t_2)
F_i(t) F_j(t-\tau) F_l(t_1) \rangle _R )\langle Q_m(t_2) Q_l(t_1)
\rangle
 _S].$\hfill(18)
\end{flushleft}
It is customary to write the reservoir Hamiltonian in the following
form:
\begin{flushleft}
\hspace{1in} $\textsl{H}_R=\sum\limits_{k} {\hbar \omega_k
(b_k^\dagger b_k}+\frac{1}{2}),$\hfill(19)\\
\end{flushleft}
%\begin{flushleft}
where $\omega_k$ is the frequency of the reservoir modes. The
time-dependent reservoir operator in the IP appearing in Eq.(18) can
be expressed in the SP as
\begin{flushleft}
\hspace{1in} $F_1(t)=\sum\limits_{p}\kappa_pe^{(i/\hbar)H_R
t}b_p e^{(-i/\hbar)H_R t}$
\end{flushleft}
\begin{flushleft}
\hspace{1.4in} $=\sum\limits_{p}\kappa_pb_pe^{-{i\omega_p
t}}$,\hfill(20a)
\end{flushleft}
\begin{flushleft}
\hspace{1in} $F_2(t)=\sum\limits_{q}\kappa_q e^{(i/\hbar)H_R
t}b_q^\dag e^{(-i/\hbar)H_R t}$
\end{flushleft}
\begin{flushleft}
\hspace{1.4in} $=\sum\limits_{q}\kappa_q b_q^\dag e^{i\omega_q
t}$,\hfill(20b)
\end{flushleft}
where $\omega_l$ ($l=p, q$) is the angular frequency of the
reservoir oscillators mode and $\kappa_l$ the coupling constants,
respectively. From Eq.(18) we now proceed to evaluate the spectral
density function $\widehat{W}_{12}^+[t] $ for $i=1$ and $j=2$,
%\pagebreak
\begin{flushleft}
\hspace{.5in}
$\widehat{W}_{12}^+[t]=-\sum\limits_{l,m}\int\limits_0^{t} d\tau
e^{i\omega_i^S\tau}{\int\limits_0^{t-\tau} {dt_1 } }
\int\limits_0^{t_1 } {dt_2 }$
\end{flushleft}
\begin{flushleft}
\hspace{.5in} \vspace{.4cm} $[(\langle F_1(t) F_2(t-\tau) F_l(t_1)
F_m(t_2) \rangle _R- \langle F_l(t_1) F_1(t) F_2(t-\tau) F_m(t_2)
\rangle _R )\langle Q_l(t_1) Q_m(t_2) \rangle _S  + $ \\
\end{flushleft}
\begin{flushleft}
\hspace{.5in} \vspace{.4cm} $(\langle F_m(t_2) F_l(t_1)
F_1(t)F_2(t-\tau) \rangle _R  - \langle F_m(t_2) F_1(t)
F_2(t-\tau) F_l(t_1) \rangle _R )\langle Q_m(t_2)Q_l(t_1)
\rangle_S],$\hfill(21)
\end{flushleft}
where, once again, the term linear in the system operators is
dropped by the symmetry argument. Throughout the treatment we assume
that the reservoir is in a thermal distribution and thus only the
diagonal terms will survive. Substituting Eqs.(13), (20a) and (20b)
in Eq.(21), we obtain (for details, see Appendix)
\begin{flushleft}
\hspace {.5in}
 $\widehat{W}_{12}^+[t] =
%\langle F_1 F_2 \rangle _R  +
%\frac{1}{{\hbar^2} }
\sum\limits_{r,s}|\kappa_r|^2 |\kappa_s |^2 e^{i(\omega
_r-\omega_s)t} \int\limits_0^{t}
{d\tau}e^{i(\omega_0-\omega_r)\tau}\int\limits_0^{t-\tau} {dt_1 }
e^{i(\omega _0 -\omega_r)t_1}$
\end{flushleft}
\begin{flushleft}
\hspace {.75in} $ \int\limits_0^{t_1}{dt_2}e^{i(\omega _s
-\omega_0)t_2}((2+\bar{n}_{R}(\omega_r)+\bar{n}_{R}(\omega_s,T_R))\langle
Q_1Q_2\rangle_S-(\bar{n}_R(\omega_r,T_R)+\bar{n}_R(\omega_s,T_R))\langle
Q_2 Q_1\rangle_S),$\hfill(22)
\end{flushleft}
where $\bar{n}_R(\omega_s,T_R)$ is the average thermal photon number
of the reservoir and we have considered the system frequency to be
$\omega_1^S = -\omega_2^S =\omega_0$ for convenience without losing
generality. Thus we note that in Eq.(22), the evolution of the
reservoir arises due to the correlation functions of the system
operators. Finally converting the sum over modes into the frequency
space integrals, we find
\begin{flushleft}
\hspace {.5in}
%\begin{array}{l}
$\widehat{W}_{12}^+[t] = \int\limits_0^\infty {d\omega _r }
\textsl{D}(\omega _r )|\kappa(\omega _r)|^2\int\limits_0^\infty
{d\omega _s } \textsl{D}(\omega _s )|\kappa(\omega _s )|^2
e^{i(\omega _r-\omega_s)t} \int\limits_0^{t}
{d\tau}e^{i(\omega_0-\omega_r)\tau}$
\end{flushleft}
\begin{flushleft}
\hspace {1in}
 $\int\limits_0^{t-\tau} {dt_1 }
e^{i(\omega _0 -\omega_r)t_1} \int\limits_0^{t_1 } {dt_2
}e^{i(\omega _s
-\omega_0)t_2}((2+\bar{n}_R(\omega_r,T_R)+\bar{n}_R(\omega_s,T_R))\langle
Q_1 Q_2\rangle_S-$
\end{flushleft}
\begin{flushleft}
\hspace {1in}
$(\bar{n}_R(\omega_r,T_R)+\bar{n}_R(\omega_s,T_R))\langle
Q_2 Q_1\rangle_S),$\hfill(23)
% \end{array}
\end{flushleft}
where $D(\omega_r)$ and $D(\omega _s)$ be the density of states
respectively. Proceeding in the similar way we can show that
$\widehat{W}_{12}^+[t]=\widehat{W}_{21}^-[t]$. The time development
of the reservoir for any simple system may be calculated from
Eq.(23). \vfill \pagebreak
\begin{center}
{\bf III. Application to simple harmonic oscillator}
\end{center}

The vibrational cooling may be modeled by a harmonic oscillator
interacting with large number of the reservoir modes resulting
damping. The free Hamiltonian and interaction term of such composite
system are given by
\begin{flushleft}
\hspace{1in} $\textsl{H}_S=\hbar\omega_0(a^\dag
a+\frac{1}{2})$\hfill(24a)
\end{flushleft}
\begin{flushleft}
\hspace{1in} $V =\hbar\sum\limits_{k}(\kappa_ka^\dag
b_k+\kappa_k^*a b_k^\dag),$\hfill(24b)
\end{flushleft}
respectively, with the generic reservoir Hamiltonian given by
Eq.(19). In Eqs.(24a) and (24b), the system operators are in the
Schr$\ddot{o}$dinger picture i.e, $Q_1=a^\dagger$ and $Q_2=a$.
Taking $\bar{n}_S(t)=\langle{a}^\dag a(t)\rangle_S$ to be the
average photon number of the system in time t and the upper limits
of integration as $t$,  $t-\tau, t_1\rightarrow \infty$, Eq.(23)
becomes
\begin{flushleft}
\hspace {.5in} $\widehat{W}_{12}^+[t] =
2\pi^2\textsl{D}^2(\omega _0 )|\kappa(\omega _0)|^4
 (\bar{n}_S(t)-\bar{n}_R(T_R)) t,
$\hfill(25)
% \end{array}
\end{flushleft}
the difference between the the instantaneous average excitation
number of the system and the thermal average photon number of the
reservoir. In deriving Eq.(25) we have neglected the principal parts
which correspond to a small Lamb shift. Substituting Eq.(25) in
Eq.(17) (with $i=1$ and $j=2$) and plucking back the resulting
equation in Eqs.(15a) and (15b), we obtain a generalized Born-Markov
master equation of the damped harmonic oscillator,
\begin{flushleft}
\hspace{.5in} $\frac{dS}{dt}=-\frac{i}{\hbar}[H_S,S]-
%\int\limits_0^t {dt'}
\frac{\gamma_1(t)}{2}[a^\dagger a S-2 a S a^\dagger +
S a^\dagger a]-\frac{\gamma_2(t)}{2}[a a^\dagger S-2
a^\dagger S a + S a a^\dagger],$\hfill(26)
\end{flushleft}
where
\begin{flushleft}
\hspace{.5in}
$\gamma_{1}(t)=\gamma(1+\bar{n}_R(T_R))+\gamma^2
(\bar{n}_S(t)-\bar{n}_R(T_R))
t,$\hfill(27a)
\end{flushleft}
\begin{flushleft}
\hspace{.5in} $\gamma_{2}(t)=\gamma\bar{n}_R(T_R)+\gamma^2
(\bar{n}_S(t)-\bar{n}_R(T_R)) t,$\hfill(27b)
\end{flushleft}
with $\gamma=2\pi|\kappa(\omega_0)|^2\textsl{D}(\omega_0)$ the decay
constant. Thus, a linear time-dependent term appearing beyond the
leading order of the decay constant becomes important if
$|(\bar{n}_S(t)-\bar{n}_R(T_R)) |>>0$. If the initial average energy
of the system is much more than the thermal average excitation of
the bath, then the time-dependent decay rate $\gamma_1(t)$ and
$\gamma_2(t)$ appreciably affect the decay dynamics in an early
time.
\par
To address the notion of thermalization in our scenario, we need to
calculate the time evolution of $\bar{n}_S(t)$ ($=\langle{a}^\dag
a(t)\rangle_S)$ from Eq.(26), which is governed by the rate equation
\begin{flushleft}
\hspace{.5in}
$\displaystyle -\frac{d\bar{n}_S(t)}{dt}=\gamma(\bar{n}_S(t)
 -\bar{n}_R(T_R))(1+\gamma t)
$\hfill (28)
\end{flushleft}
and its solution reads
\begin{flushleft}
\hspace{.5in} $\displaystyle \bar{n}_S(t) =
\bar{n}_R(T_R) +(\bar{n}_S(0)- \bar{n}_R(T_R))
\displaystyle e^{-\gamma t(1+\frac{\gamma t}{2})}.$ \hfill (29)
\end{flushleft}
Comparing Eq.(5) with Eq.(29), we note that in the later case the
decay rate is time-dependent and thermalization becomes faster due
to the incorporation of the dynamical evolution of the reservoir.
\par
Here, we note that in the high temperature approximation, Eq.(28)
leads to the modified Newton's law of cooling as
\begin{flushleft}
\hspace{.5in}
$-\frac{dT(t)}{dt}=\gamma(T(t)-T_R)+\gamma^2(T(t)-T_R)t,\quad
$\hfill (30)
\end{flushleft}
where the term beyond the leading order of the decay constant
becomes significant if $T_0>>T_R$. The Fig. shows the comparison of
the plots of Eq.(2) with Eq.(30), where we note that the
thermalization occurs at a faster rate. Cooling in early time is
much faster than predicted by Newton's law. As a first order
correction the theory is valid upto $t<\gamma^{-1}$; for a time
longer than $t \ge \gamma^{-1}$, Newton's exponential law should be
considered for thermalization. Cooling-time for reaching from
$2000^{0}C$ to $800^{0}C$ is almost two-third in the modified
dynamics of that in the Newton's cooling law. In comparison to
Newton's law, where the time required to bring the temperature of
the system above the reservoir-temperature to its half, i.e.,
$\displaystyle \frac{T(t)-T_R}{T(0)-T_R}= \frac{1}{2}$,
half-thermalization time is, $t_{1/2} = \frac{\ln{2}}{\gamma}$,
whereas for the modified case the half-thermalization time is given
by $t_{1/2} = \frac{\sqrt{1+2\ln{2}}-1}{\gamma}$. Therefore, in the
modified case also $t_{1/2}$ is independent of $(T(0)-T_R)$.

\begin{figure}
%\centerline{ \epsffile{Figure.eps}}
\centerline{
%\rotatebox{-90}{\resizebox{4.0in}{4.0in}{\includegraphics{Figure.ps}}}}
\rotatebox{-90}{\includegraphics[width=3.5in]{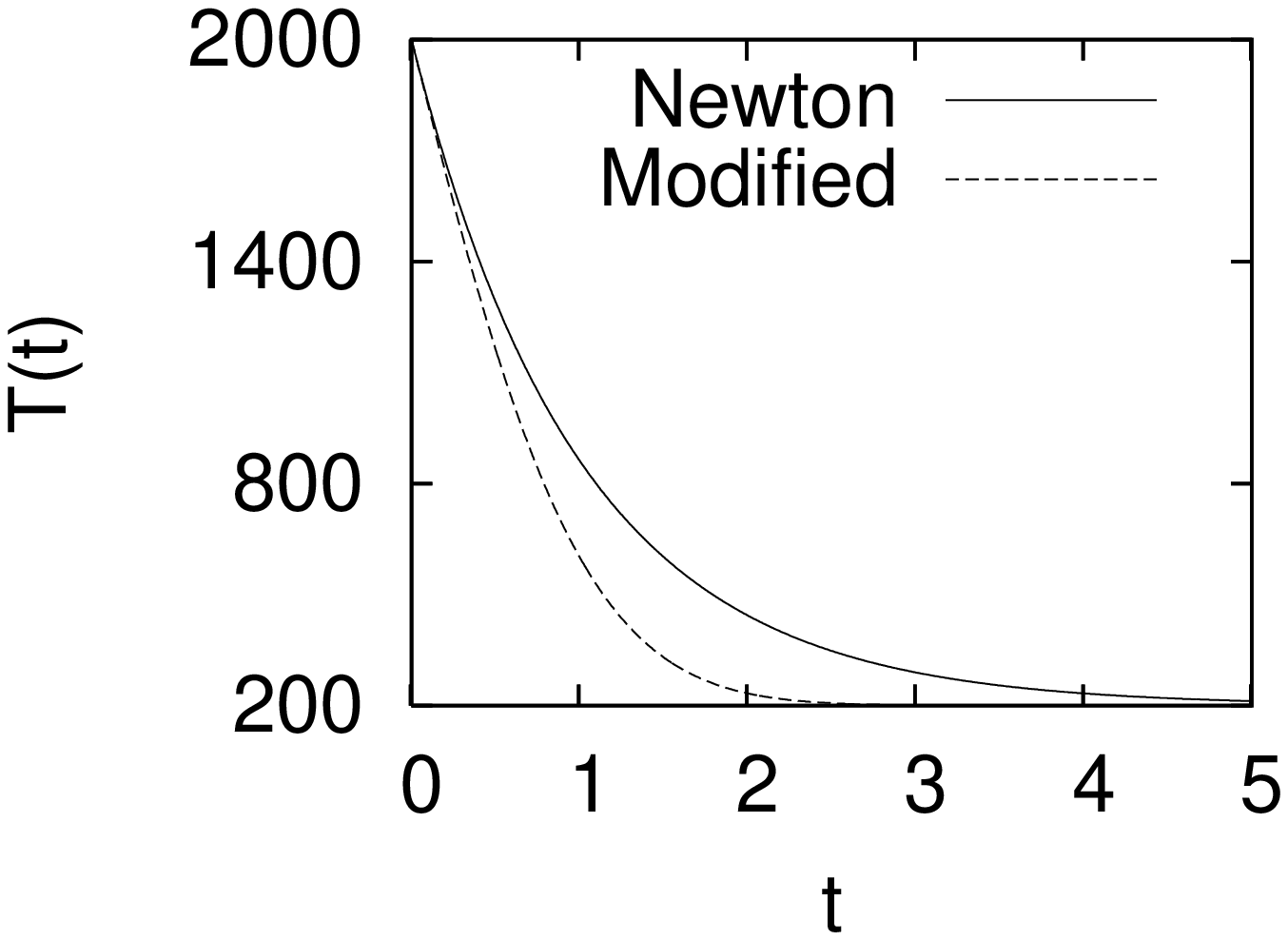}}}
\noindent \small \bf {[Fig.]: The plot of Newton's cooling law is
given by Eq.(2) (Newton, solid line) and its modified form given by
Eq.(30) (Modified, dashed line) with time(in $\gamma^{-1}$) for
initial and final temperatures(units arbitrary), $T_0=2000^0 C$ and
$T_R=200^0 C$.}
\end{figure}
\par
From Eq.(26), the corresponding master equation of the diagonal
elements of the density matrix can be given by the loss-gain
equation of population for the step-ladder model of a harmonic
oscillator as

\begin{flushleft}
\hspace{.5in} $\displaystyle \frac{dS_{ii}(t)}{dt} = \sum_{j=i\pm
1}^{\infty}
 [ W(j|i) S_{jj}(t) - W(i|j) S_{ii}(t)].$\hfill(31)
\end{flushleft}
Here the transition rates connecting only the neighbouring levels
are specifically given by
\begin{flushleft}
\hspace{.5in} $ W(i+1|i) = (i+1) \gamma[1+\{\bar{n}_{R} +
(\bar{n}_S(t)-\bar{n}_{R})
\gamma t\}]$\\
\hspace{.5in} $W(i|i+1) = i \gamma[\bar{n}_{R} +
(\bar{n}_S(t)-\bar{n}_{R}) \gamma t],$ \hfill(32)
\end{flushleft}
which means that the $(i+1)$-th to $i$-th state transition rate is
time dependent, and vice-versa. On the top of the usual
temperature-dependent rate, it depends on $\gamma t$ and on the
difference in temperature of the reservoir from that of the
instantaneous temperature of the system, $T_S(t)-T_R$ (or
equivalently $(\bar{n}_S(t)-\bar{n}_{R})$). Usually we consider
$\gamma t <1$ for the first-order perturbative effect. A direct
consequence of the loss-gain-type master equation shows that the
bath-induced forward and backward rates are modified by a factor of
$\bar{n}_{R} + (\bar{n}_S(t)-\bar{n}_{R}) \gamma t$ instead of
$\bar{n}_R$. As the modification arises due to the system-induced
dynamics of the reservoir, which is considered as a first-order
effect, we can safely assume that the rate is primarily governed by
the factor $\bar{n}_R$ and therefore we can consider
$(\bar{n}_S(t)-\bar{n}_{R}) \gamma t \le \bar{n}_R$. This amounts to
the fact that the initial system temperature should not be
arbitrarily high compared to the temperature of the reservoir.
Otherwise, a strong non-equilibrium evolution of the bath will
produce a nonlinear coupled dynamics of system and bath variables,
which is immensely difficult problem to tackle to provide any
tangible physical result.
%\pagebreak
\begin{center}
\bf{IV. Conclusion}
\end{center}
\par
In this paper we have developed the quantum theory of cooling of a
system with large energy content when the reservoir has also a
dynamical evolution rather than thermal equilibrium. It is
explicitly demonstrated for such system
%within the framework of the damped harmonic oscillator that
 that the thermal equilibrium is attained much faster than in comparison to
 the case of exponential decay when the reservoir is at equilibrium.
Our study reveals that, the larger is the initial photon number
content of the system, the faster is the rate of cooling. Possible
modification of the Newton's classical law of cooling beyond the
leading order of the decay constant is pointed out and it is shown
that the higher order term becomes significant if the difference
between the average energy per mode of the system and the reservoir
is considerable. The analysis is strictly valid for a very short
time, $t<\gamma^{-1}$, and initial temperature or average energy of
the system is not too much higher compared to the reservoir as the
modification in the theory arises due to first order perturbation
effect. We have also considered a repeated neglect of off-diagonal
terms corresponding to bath degrees of freedom arising from the
dynamics where only the diagonal elements of the bath density are
modified in time. A faster thermalization or faster cooling is
qualitatively understandable as the bath is interacting with the
system more actively instead of passively waiting in its equilibrium
distribution in the usual approach.
\par
To treat a finite size [18,19] of the bath, one immediate choice is
to restrict the number of modes in the bath. This is equivalent to a
pronounced recurrence of population of the vibrational states due to
the back and forth exchange of energy between the modes of the
system and bath. However, in our approach we have assumed the fact
that the system experiences a feedback due to the dynamical
evolution of the bath, but ultimately the bath is assumed to be in
thermal equilibrium. We have calculated the two-point and four-point
correlation functions of the bath variables by averaging over the
thermal equilibrium distribution instead of a non-equilibrium
distribution of the bath. The population decay is non-exponential
due to this, which has simple dependence on the difference between
the average energy of each mode from the time-dependent state of the
system to the bath at equilibrium. A more systematic approach to
treat finite size of the bath is under consideration which will be
published elsewhere.
\par
In the midst of several currently interesting coherent cooling
mechanism of atoms and molecules induced by laser, this incoherent
mechanism of vibrational cooling may find it worthwhile because of
the huge availability of nano-materials [20] which can support a
large number of degrees of freedom. It can effectively act as a bath
as well as a finite quantum system to reciprocates energy with a
subsystem of interest which is composed of a single or a few modes
of vibration. Other associated aspects of the system-reservoir
formalism require careful scrutiny in the light of the dynamical
evolution of the reservoir proposed here. \vfill
\begin{center}
\large Acknowledgement
\end{center}
\par
MRN thanks the University Grants Commission, New Delhi and SS thanks
Department of Science and Technology, New Delhi for financial
support. We thank Professor D S Ray for many fruitful discussions.
MRN and SS also thank to Dr A K Sen for his interest in this problem
and acknowledge the hospitality of S N Bose National Centre for
Basic Sciences, Kolkata, where part of the work was carried out.
\pagebreak
%and acknowledge the Associateship of S N Bose National Centre for
%Basic Sciences, Kolkata. \pagebreak
\begin{center}
Appendix
\end{center}
\par
In this appendix we shall derive the four-point reservoir
correlation functions appearing in Eq.(21). The four-point reservoir
correlators can be expressed in terms of two-point correlators by
using the identity,
\begin{flushleft}
\hspace{.5in}
 $\langle
\textsl{O}_a\textsl{O}_b\textsl{O}_c\textsl{O}_d\rangle_R= \langle
\textsl{O}_a\textsl{O}_b\rangle_R \langle
\textsl{O}_c\textsl{O}_d\rangle_R+ \langle
\textsl{O}_a\textsl{O}_c\rangle_R \langle
\textsl{O}_b\textsl{O}_d\rangle_R+\langle
\textsl{O}_a\textsl{O}_d\rangle_R\langle
\textsl{O}_b\textsl{O}_c\rangle_R$.\hfill (A.1)
\end{flushleft}
Using Eq.(A.1) let us calculate the term in the square bracket of
Eq.(21) with $i=1$, $j=2$ and $l,m$ runs from 1 to 2,
\begin{flushleft}
\hspace{.5in}
$\sum_{l,m}[\langle\textsl{F}_1(t)\textsl{F}_2(t')\textsl{F}_l(t_1)\textsl{F}_m(t_2)\rangle_R-
\langle\textsl{F}_l(t_1)\textsl{F}_1(t)\textsl{F}_2(t')\textsl{F}_m(t_2)\rangle_R]\langle{Q_lQ_m}
\rangle_S$
\end{flushleft}
\begin{flushleft}
\hspace{.5in}
$+[\langle\textsl{F}_m(t_2)\textsl{F}_l(t_1)\textsl{F}_1(t)\textsl{F}_2(t')\rangle_R-
\langle\textsl{F}_m(t_2)\textsl{F}_1(t)\textsl{F}_2(t')\textsl{F}_l(t_1)\rangle_R]\langle{Q_mQ_l}\rangle_S
=$
\end{flushleft}
\begin{flushleft}
\hspace{1in} $[\langle\textsl{F}_1(t_1)\textsl{F}_2(t')\rangle_R
(\langle\textsl{F}_2(t_2)\textsl{F}_1(t)\rangle_R-\langle\textsl{F}_1(t)\textsl{F}_2(t_2)\rangle_R)+$
\end{flushleft}
\begin{flushleft}
\hspace{1in} $\langle\textsl{F}_1(t)\textsl{F}_2(t_2)\rangle_R
(\langle\textsl{F}_2(t')\textsl{F}_1(t_1)\rangle_R-\langle\textsl{F}_1(t_1)\textsl{F}_2(t')\rangle_R)]\langle{Q_1Q_2}\rangle_S$
\end{flushleft}
\begin{flushleft}
\hspace{1in} $[\langle\textsl{F}_2(t')\textsl{F}_1(t_1)\rangle_R
(\langle\textsl{F}_1(t)\textsl{F}_2(t_2)\rangle_R-\langle\textsl{F}_2(t_2)\textsl{F}_1(t)\rangle_R)+$
\end{flushleft}
\begin{flushleft}
\hspace{1in} $\langle\textsl{F}_2(t_2)\textsl{F}_1(t)\rangle_R
(\langle\textsl{F}_1(t_1)\textsl{F}_2(t')\rangle_R-\langle\textsl{F}_2(t')\textsl{F}_1(t_1)\rangle_R)]\langle{Q_2Q_1}\rangle_S.$\hfill
(A.2)
\end{flushleft}
%\hspace{.5in}
To obtain the above equation we have neglected the off-diagonal
terms, since the reservoir assumed to be a thermal one. Substituting
Eqs.(13), (20a) and (20b) in Eq.(A.2), the straightforward
simplification yields
\begin{flushleft}
\hspace{.5in}
$\sum_{l,m}[\langle\textsl{F}_1(t)\textsl{F}_2(t')\textsl{F}_l(t_1)\textsl{F}_m(t_2)\rangle_R-
\langle\textsl{F}_l(t_1)\textsl{F}_1(t)\textsl{F}_2(t')\textsl{F}_m(t_2)\rangle_R]\langle{Q_lQ_m}\rangle_S$
\end{flushleft}
\begin{flushleft}
\hspace{.5in}
$+[\langle\textsl{F}_m(t_2)\textsl{F}_l(t_1)\textsl{F}_1(t)\textsl{F}_2(t')\rangle_R-
\langle\textsl{F}_m(t_2)\textsl{F}_1(t)\textsl{F}_2(t')\textsl{F}_l(t_1)\rangle_R]\langle{Q_mQ_l}\rangle_S
=$
\end{flushleft}
\begin{flushleft}
\hspace {1in} $-\sum\limits_{r,s}|\kappa_r|^2 |\kappa_s |^2
e^{i(\omega _r-\omega_s)t} e^{-i(\omega_r t_1-\omega_s t_2)}
e^{-i\omega_r \tau}$
\end{flushleft}
\begin{flushleft}
\hspace {1in} $((2+\bar{n}_R(\omega_r)+\bar{n}_R(\omega_s))\langle
Q_1 Q_2\rangle_S-(\bar{n}_R(\omega_r)+\bar{n}_R(\omega_s))\langle
Q_2 Q_1\rangle_S ),$\hfill(A.3)\\
\end{flushleft}
where $t'$ is replaced by $t-\tau$. Finally substituting Eq.(A.3)
in Eq.(21) we obtain Eq.(22). \pagebreak


\begin{thebibliography}{5.8 in}
\bibitem [1] {CP}  C  Cohen-Tannoudji and W D Phillips, Phys. Today
{\bf 43},
33 (1990)
\bibitem [2] {Cohen} C  Cohen-Tannoudji, Phys. Rep. {\bf 219}, 153
(1992) and references therein.
\bibitem [3] {Gabbinini} C Gabbinini and A Fioretti in Trapped
Particles and Fundamental Physics, Ed S N Atutov, R Calabrese and L Moi,
(Kauwar Academic Press, London, 2002) pp181 and references therein.
\bibitem[4] {Louisell} W H Louisell, Quantum Statistical Properties of
Radiation
(Wiley, New York,1973) pp 336
\bibitem[5] {Orszag} M Orszag, Quantum Optics (Springer,
Heidelberg, 2000) pp 9
\bibitem[6] {lin}  A. A. Villaeys, J. C. Vallet and S. H. Lin, Phys.
Rev. A 43, 5030 (1991)
\bibitem [7] {van} N G van Kampen Phys. Rep. {\bf 24}, 171 (1976)
\bibitem [8] {Mukamel1} S Mukamel, I Oppenheim and J Ross, Phys. Rev.
{\bf
A17}, 1988 (1978)
\bibitem [9] {ullersma} P Ullersma,Physica {\bf 32},  27, 56, 74, 90
(1996)
\bibitem [10] {gango1} G Gangopadhyay and D S Ray, Phys. Rev. {\bf
A44}, 2206 (1991)
\bibitem [11] {gango2} G Gangopadhyay and D S Ray, Phys. Rev. {\bf
A46}, 1507 (1992)
\bibitem [12] {gango3} G Gangopadhyay and S Ghosal, Chem Phys Lett {\bf
289}, 287 (1999)
\bibitem [13] {Brecker} P C Brecker, H L Fragnito, J Y Bigot, C H
Britocruz, C V
Shank, Phys, Rev. Lett. {\bf 63}, 505 (1973)
\bibitem [14] {vogel} W Vogel, D G Welsch, W Wilhelmi, Phys. Rev. {\bf
A37}, 3825 (1988)
\bibitem [15] {Tchenio} P Tchenio, A Debarre, J C Keller, J L Le Gouet,
Phys Rev Lett {\bf 62}, 415 (1989)
\bibitem [16] {Winters} M P Winters and P E Toschek, Phys Rev Lett {\bf
65}, 3116 (1990)
\bibitem [17] {Walther} H Walther, Phys Rep {\bf 219}, 263 (1990)
\bibitem [18] {Andrianov} I Andrianov and P Saalfrank, Chem Phys
Lett {\bf{433}}, 91 (2006)
\bibitem [19] {Beck} M H Beck, A Jackle, G A Worth, and H D Meyer,
Phys. Rep. {\bf{324}}, 1 (2000)
\bibitem [20] {CAO} See, for example, G Cao, Nanostructures and
Nanomaterials: Synthesis, Properties and Application(Imperial
College Press, London, 2004) and references therein.
\end{thebibliography}
\end{document}